\documentclass[12pt]{iopart}
\usepackage{amssymb}
\usepackage{iopams}

\begin{document}

\title[Collective mode description]{Collective mode description for a clean
Fermi gas with repulsion in arbitrary dimensions: interaction of
spin excitations and its consequences}
\author{K.B. Efetov${}^{1,2}$, I.L. Aleiner$^3$}

\address{$^1$ Theoretische Physik III,
Ruhr-Universit\"{a}t Bochum, 44780 Bochum, Germany}
\address{$^2$ L.D. Landau Institute for Theoretical Physics,
117940 Moscow, Russia}
\address{$^3$ Physics Department, Columbia
University, New York, NY 10027, USA}

\begin{abstract}
We discuss a recent bosonization method developed to study clean
Fermi gases with repulsion in any dimensions. The method enables
one to consider both density and spin excitations. It is
demonstrated that due to a non-abelian structure of the effective
theory, the spin excitations interact with each other, which leads
to new logarithmic in temperature corrections to physical
quantities. Using a renormalization group scheme constructed for
the effective low energy field theory these logarithms are summed
up in all orders. Temperature dependent corrections to the
specific heat and spin susceptibility are obtained for all
dimensions $d=1,2,3.$
\end{abstract}

\maketitle

\section{Introduction}

Landau theory of the Fermi liquid (FL) \cite{landau} is by now one of the
most established theories describing a system of interacting fermions. The
main statement of this theory is that the low energy behavior of the system
is similar to that for the ideal Fermi gas. This gives a possibility to
discuss experimental systems omitting the interactions and using
phenomenological constants for the effective mass of the particles, density
of states at the Fermi surface and other physical quantities.

A phenomenological description of the FL developed in the first works \cite%
{landau} was supported by a diagrammatic analysis \cite{landau1}, which was
a very good confirmation. However, the microscopic Landau theory of Ref.
\cite{landau1} is based on a very strong assumption that one can single out
a singular particle-hole channel and sum proper ladder diagrams. Irreducible
vertices entering the ladder diagrams should remain finite and be analytic
in the limit of small momenta and frequencies. It is generally believed that
for a Fermi gas with repulsion this assumption is correct and such a system
should behave as the Fermi liquid in dimensions $d>1$.

Of course, the similarity between the FL and ideal Fermi gas cannot be exact
and there should be corrections at finite temperatures, frequencies or
momenta. Study of unconventional metals like high temperature
superconductors, heavy fermion materials, etc., have revealed considerable
deviations of their properties from those predicted by the FL. As a result,
quite a few theoretical works have appeared recently where the Landau FL
theory was discussed \ \cite{anderson,shankar,metzner,houghton,abanov}.

The corrections to physical quantities become especially interesting when
they are non-analytic functions of temperature, frequency or momentum. This
means that physical quantities like $C\left( T\right) /T,$ where $C\left(
T\right) $ is the specific heat and $T$ is the temperature, $\chi \left(
T\right) $, where $\chi \left( T\right) $ is the spin susceptibility, etc.,
cannot be represented at low temperatures as a series in $T^{2}$, which
contrasts the ideal Fermi gas. Such non-analytical corrections were studied
in a number of publications using diagrammatic expansions in the
electron-electron interaction \ \cite%
{eliash,doniach,brink,amit,belitz,coffey,chubukov1,sarma,chubukov2,chubukov3,chubukov4}%
. As a result of this investigation, it is well known by now that in $d=3$
the next-to-leading term in $C\left( T\right) /T$ is proportional to $%
T^{2}\ln T$ \cite{eliash,doniach,brink,amit}. In $2d,$ the non-analytical
corrections to $C\left( T\right) /T$ and $\chi \left( T,Q\right) $ scale as $%
T$ and $\max \left\{ T,v_{0}Q\right\} $, respectively ($Q$ is the wave
vector of the external magnetic field) \cite%
{coffey,chubukov1,sarma,chubukov2,chubukov3,chubukov4,catelani}.

The existence of the non-analytic corrections to the physical quantities is
not accidental. In fact, all the singular corrections to the thermodynamic
quantities may be understood in terms of contributions of low lying
collective excitations, see e.g., Refs. \cite{catelani,chubukov3}. At the
same time, explicit calculations with the conventional diagrammatic
technique are not simple already in the lowest orders of the perturbation
theory. Therefore, selecting diagrams in order to group them into the
collective modes is a rather difficult task.

In this paper we present a new method of calculations for a clean Fermi gas
with a repulsion that enables us to \textquotedblleft integrate
out\textquotedblright\ electron degrees of freedom in the beginning of all
calculations and reduce the initial fermionic model to a model of low lying
excitations. The method is based on using equations for quasiclassical Green
functions and includes both the density and spin excitations. We call
loosely our method \emph{bosonization }but it differs from earlier higher
dimensional bosonization schemes \cite%
{haldane,houghton1,houghton,neto,kopietz,kopietzb,khveshchenko,khveshchenko1,metzner}
based on the assumption a long range electron-electron interaction. As a
result, only density excitations were considered, while the spin excitations
were not included in those schemes (to be more precise, the spin excitations
are not affected by the long range interaction and therefore they were not
included to the scheme).

The density excitations are described by a scalar function and the effective
interaction $V_{c}\left( k,\omega \right) $ between them vanishes in the
limit $\omega ,k\rightarrow 0.$ This means that the interaction may lead
only to a renormalization of parameters characterizing physical quantities
(unless \ $V_{c}\left( k,\omega \right) $ is long ranged) but not to new
effects. In other words, the bosonization of \cite%
{haldane,houghton1,houghton,neto,kopietz,kopietzb,khveshchenko,khveshchenko1,metzner}
enables one to reduce a system with a long range interaction to a model of
free bosons, which is very similar to writing the Tomonaga-Luttinger model
for one dimensional electron systems (see e.g. \cite{tsvelik}). This method
is often referred to as bosonization.

What we want to present now is a scheme that enables one to consider for
arbitrary interactions both the density and spin excitations on equal
footing. The relevant variable describing the spin excitations is a $2\times
2$ matrix and our method resembles to some extent a non-abelian
bosonization. The effective interaction between the spin excitations does
not vanish in the limit $\omega ,k\rightarrow 0$ and leads to new
logarithmic contributions to scattering amplitudes coming from low energies
of the order of $T$. As a result, physical quantities depend on the
logarithmically renormalized amplitudes, which changes their temperature
dependence. It is important to emphasize that the \textquotedblleft
infrared\textquotedblright\ logarithmic divergencies that we find in the
limit $T\rightarrow 0$ exist in \emph{any }dimensions (including $d=1$) and
have nothing in common with the \textquotedblleft
ultraviolet\textquotedblright\ logarithmic divergency (divergency
originating from short distances) discussed for $2d$ systems long ago \cite%
{galitskii}. The latter does not lead to any additional dependence on
temperature and can be absorbed into parameters characterizing FL. \emph{\ }

We display in the subsequent Sections the main idea of our method and new
results that have been obtained recently. In Sec.\ref{excitations} we make a
Hubbard-Stratonovich transformation and derive quasiclassical equations for
the density and spin excitations representing their solution in terms of an
integral over supervectors. The interaction between the spin excitations
gives logarithms that are summed in Sec.\ref{field theory} using a new
renormalization group (RG) scheme. We calculate the specific heat and spin
susceptibility in Sec.\ref{thermodynamics} and discuss the results in Sec.%
\ref{discussion}.

The method and the calculation of the specific heat has been presented for
the first time in Ref. \cite{aleiner}. Using this method the calculation of
the spin susceptibility was carried out later in Ref. \cite{schwiete}. Our
presentation here is based on these publications but we concentrate rather
on explaining the main steps of the derivation than on explicit
calculations. All necessary details can be found in Refs. \cite%
{aleiner,schwiete}.

\section{Spin and density excitations and their contribution to
thermodynamics.}

\label{excitations}

\subsection{Singling out slow pairs and Hubbard-Stratonovich transformation.}

In this Section we show how one can reduce calculation of the partition
function of the interacting fermions to calculation of the partition
function of the density and spin excitations. It will be demonstrated that
the density excitations are described by a model of free bosons, whereas
spin excitations interact with each other.

We start the discussion writing the original partition function $Z$ in terms
of a functional integral over classical anticommuting variables $\chi
_{\sigma }\left( x\right) $ ($x=\left\{ \mathbf{r,}\tau \right\} $ and $%
\sigma $ labels the spin)%
\begin{equation}
Z=\int \exp \left( -\mathcal{S}\right) D\chi D\chi ^{\ast }  \label{a1a}
\end{equation}%
The action $\mathcal{S}$ entering Eq. (\ref{a1a}) has the form
\begin{equation}
\mathcal{S=S}_{0}+\mathcal{S}_{int},  \label{a2}
\end{equation}%
where the term $\mathcal{S}_{0},$
\begin{equation}
\mathcal{S}_{0}=\sum_{\sigma }\int \chi _{\sigma }^{\ast }\left( x\right)
\left( -\frac{\partial }{\partial \tau }-\frac{\mathbf{\hat{p}}^{2}}{2m}%
+\epsilon _{F}\right) \chi _{\sigma }\left( x\right) dx  \label{a3}
\end{equation}%
stands for the action of free fermions ($\epsilon _{F}$ is the Fermi energy,
$m$ is the mass and $\mathbf{\hat{p}}$ is the momentum operator). Eqs. (\ref%
{a2}, \ref{a3}) are written in the Matsubara representation with the
imaginary time $\tau ,$ such that the field variables $\chi \left( \mathbf{r,%
}\tau \right) $ are antiperiodic in $\tau $%
\begin{equation}
\chi \left( \mathbf{r,}\tau \right) =-\chi \left( \mathbf{r},\tau +1/T\right)
\label{a5}
\end{equation}%
The term $\mathcal{S}_{int}$ is Eq. (\ref{a2}) describes the fermion-fermion
interaction,
\begin{equation}
\mathcal{S}_{int}=\frac{1}{2}\sum_{\sigma ,\sigma ^{\prime }}\int \chi
_{\sigma }^{\ast }\left( x\right) \chi _{\sigma ^{\prime }}^{\ast }\left(
x^{\prime }\right) v\left( x-x^{\prime }\right) \chi _{\sigma ^{\prime
}}\left( x^{\prime }\right) \chi _{\sigma }\left( x\right) dxdx^{\prime },
\label{a5a}
\end{equation}%
where $v\left( x-x^{\prime }\right) =V\left( \mathbf{r-r}^{\prime }\right)
\delta \left( \tau -\tau ^{\prime }\right) $ and $V\left( \mathbf{r-r}%
^{\prime }\right) $ is the potential of the interaction.

Presence of a magnetic field $\mathbf{b}$ acting on spin can be accounted
for by adding an addition term $\mathcal{S}_{b}$ to the action%
\begin{equation}
\mathcal{S}_{b}=\int dx\;\chi _{\sigma }^{\ast }(x)\mathbf{b\sigma }_{\sigma
\sigma ^{\prime }}\chi _{\sigma ^{\prime }}(x),  \label{a4a}
\end{equation}
Inclusion of this term is necessary for calculation of the spin
susceptibility $\chi \left( T\right) $.

The functional integral over $\chi _{\sigma }\left( x\right) $ in Eq. (\ref%
{a1a}) is too complicated to be calculated exactly and we restrict ourselves
with the case of a weak interaction. A stronger interaction may renormalize
the coupling constants but does not seem to change the temperature behavior.

In order to reduce the fermionic model to the model of the low lying
excitations we single out in the interaction term $\mathcal{S}_{int}$, Eq. (%
\ref{a5a}), pairs $\chi ^{\ast }\chi $ slowly varying in space and write the
effective interaction $\mathcal{S}_{int}$ as

\begin{eqnarray}
\mathcal{S}_{int} &\mathcal{\rightarrow }&\frac{1}{2}\sum_{\sigma ,\sigma
^{\prime }}\int dP_{1}dP_{2}dK\{V_{2}\chi _{\sigma }^{\ast }\left(
P_{1}\right) \chi _{\sigma }\left( P_{1}+K\right) \chi _{\sigma ^{\prime
}}^{\ast }\left( P_{2}\right) \chi _{\sigma ^{\prime }}\left( P_{2}-K\right)
\nonumber \\
&&-V_{1}\left( \mathbf{p}_{1}-\mathbf{p}_{2}\right) \chi _{\sigma }^{\ast
}\left( P_{1}\right) \chi _{\sigma ^{\prime }}\left( P_{1}+K\right) \chi
_{\sigma ^{\prime }}^{\ast }\left( P_{2}\right) \chi _{\sigma }\left(
P_{2}-K\right) \}  \label{a6}
\end{eqnarray}

In Eq. (\ref{a6}), $P_{i}=\left( \mathbf{p}_{i}\mathbf{,}\varepsilon
_{n_{i}}\right) ,$ where $\mathbf{p}_{i}$ is the momentum and $\varepsilon
_{n_{i}}=\pi T\left( 2n_{i}+1\right) $ are Matsubara fermionic frequencies $%
(i=1,2)$. As concerns $K$, it has the form $K=\left( \mathbf{k,}\omega
_{n}\right) $, where $\omega _{n}=2\pi Tn$ are Matsubara bosonic frequencies.

The symbols of the integration $\int dP_{i}$ and $\int dK$ in Eq. (\ref{a6})
read as follows
\begin{equation}
\int dP_{i}\left( ...\right) =T\sum_{\varepsilon _{n_{i}}}\frac{d^{d}\mathbf{%
p}}{\left( 2\pi \right) ^{d}}\left( ...\right) , \int dK\left( ...\right)
=T\sum_{\omega _{n}\neq 0}f\left( \mathbf{k}\right) \frac{d^{d}\mathbf{k}}{%
\left( 2\pi \right) ^{d}}\left( ...\right)  \label{a7a}
\end{equation}
where
\begin{equation}
f\left( \mathbf{k}\right) =f_{0}\left( kr_{0}\right)  \label{a7c}
\end{equation}
and $k=\left\vert \mathbf{k}\right\vert $. The function $f_{0}\left(
t\right) $ has the following asymptotics: $f_{0}\left( t\right) =1$ at $t=0$
and $f\left( t\right) \rightarrow 0$ at $t\rightarrow \infty $.

The function $f\left( \mathbf{k}\right) $ in Eq. (\ref{a7c}) is written in
order to cut off large momenta $k.$ The parameter $r_{0}$ is the minimal
length in the theory and we assume that $r_{0}\gtrsim p_{F}^{-1}.$ So, the
momenta $k$ are cut by the maximal momentum $k_{c}=r_{0}^{-1}$ and we avoid
double counting when calculating the partition function $Z$.

Introducing the cutoff $r_{0}$ means that the pairs written in Eq. (\ref{a6}%
) vary slowly in space. Accordingly, we neglect the dependence of $V_{1}$
and $V_{2}$ on the momentum $\mathbf{k}$ in Eq. (\ref{a6}). Although being
smaller than the Fermi momentum $p_{F}$ and the Fermi energy $\varepsilon
_{F}$, the cutoff $k_{c}$ is larger than all other momenta in the model.

Additional decoupling in the Cooper channel is not included, since this
would amount to overcounting of relevant scattering processes \cite{aleiner}%
. To be short, the most important is parallel or antiparallel motion of the
particles and one needs to consider only forward and backward scattering. In
this limit, adding the Cooper channel would mean double counting. This is
not so for disordered systems where all scattering angles are important and
where one should take into account the Cooper channel \cite{finkel}.

For a short range potential we can further simplify our considerations be
setting $V_{2}=V(\left\vert \mathbf{q}\right\vert \ll p_{F})$. Since
important momenta are close to the Fermi surface we can write $V_{1}(\theta
_{12})=V(\mathbf{p}_{1}-\mathbf{p}_{2})=V(2p_{0}\sin (\frac{\theta _{12}}{2}%
))$, where $\theta _{12}$ is the angle between momenta $\mathbf{p}_{1}$ and $%
\mathbf{p}_{2}$, $\theta _{12}=\widehat{\mathbf{p}_{1}\mathbf{p}_{2}}$.

For the further development of the theory it will be crucial to separate
explicitly interactions in the triplet and singlet channel. Making the
notations
\begin{equation}
V_{s}(\theta _{12})=V_{2}-\frac{1}{2}V_{1}(\theta _{12}),\quad V_{t}(\theta
_{12})=\frac{1}{2}V_{1}(\theta _{12})\qquad  \label{a7d}
\end{equation}%
one can represent the interaction term in a form of a sum of charge and a
spin parts,
\begin{eqnarray}
\tilde{\mathcal{S}}_{int} &=&\mathcal{S}_{int,s}+\mathcal{S}_{int,t}\;,
\label{a8} \\
\mathcal{S}_{int,s} &=&\frac{1}{2}\int dp_{1}dp_{2}dq\;\rho
(p_{1},-q)V_{s}(\theta _{12})\rho (p_{2},q)\;, \\
\mathcal{S}_{int,t} &=&-\frac{1}{2}\int dp_{1}dp_{2}dq\;\mathbf{S}%
(p_{1},-q)V_{t}(\theta _{12})\mathbf{S}(p_{2},q)\;,\quad
\end{eqnarray}%
where the charge $\rho (p,q)$ and spin densities $\mathbf{S}(p,q)$ are
\begin{equation}
\rho (p,q)=\chi ^{\dagger }\left( p-\frac{q}{2}\right) \chi \left( p+\frac{q%
}{2}\right) ,\quad \mathbf{S}(p,q)=\chi ^{\dagger }\left( p-\frac{q}{2}%
\right) \mathbf{\sigma }\chi \left( p+\frac{q}{2}\right) \;,  \label{a10}
\end{equation}%
and we turned to a spinor notation $\chi =(\chi _{\uparrow },\chi
_{\downarrow })$.

In order to simplify the presentation we do not write for a while the
function $f$ assuming that the variables $\rho $ and \ $S$ are not equal to
zero for small \ $q$ only, which corresponds to a slow variation of these
variables in space.

Having written the interaction term $\tilde{\mathcal{S}}_{int}$ in the form
of Eq. (\ref{a8}) we next decouple it using a Hubbard-Stratonovich
transformation with a field
\begin{equation}
\phi _{\mathbf{n}}(x)\equiv i\varphi _{\mathbf{n}}(x)+\mathbf{\sigma }{%
\mathbf{h}}_{\mathbf{n}}(x)  \label{a10a}
\end{equation}%
Here $\varphi _{\mathbf{n}}(x)$ and $\mathbf{h}_{\mathbf{n}}(x)$ are real
bosonic fields, so that $\phi _{\mathbf{n}}(\mathbf{r},\tau )=\phi _{\mathbf{%
n}}(\mathbf{r},\tau +\beta )$ and $\mathbf{n}$ is the direction of momentum $%
\mathbf{p}$ on the Fermi surface, $\mathbf{n}=\mathbf{p}/|\mathbf{p}|$. The
result is the following representation of the partition function (we omit
for a while the external field $\mathbf{b}$)%
\begin{equation}
\mathcal{Z}=\int \mathcal{D}\phi \;W_{s}[\varphi ]W_{t}[\mathbf{h}]\mathcal{Z%
}[\mathbf{h},\varphi ]\;.  \label{a11}
\end{equation}

The weight functions $W_{s}$, $W_{t}$ are shown below in Eqs.~(\ref{a13a}, %
\ref{a13b}). The partition function $\mathcal{Z}[\mathbf{h},\varphi ]$
describes the fermion motion for a fixed configuration of fields $\mathbf{h}%
,\varphi $
\begin{equation}
\mathcal{Z}[\phi ]=\int D(\chi ^{\ast },\chi )\;\exp (-S_{eff}[\phi ]).
\label{a11a}
\end{equation}%
where the effective action $\mathcal{S}_{eff}$ has the form
\begin{equation}
\mathcal{S}_{eff}[\phi ]=\mathcal{S}_{0}+\int d\mathbf{p}d\mathbf{r}_{1}d%
\mathbf{r}_{2}\chi ^{\dagger }(\mathbf{r}_{1},\tau )\phi _{\mathbf{n}}\left(
\frac{\mathbf{r}_{1}+\mathbf{r}_{2}}{2}\right) \chi (\mathbf{r}_{2},\tau
)e^{i\mathbf{p}(\mathbf{r}_{1}-\mathbf{r}_{2})}\;.\qquad  \label{a12}
\end{equation}%
Now we can write down a representation of the partition function in the
presence of the magnetic field as a weighted integral over field
configurations
\begin{equation}
\mathcal{Z}=\int \mathcal{D}\phi \;W_{s}[\varphi ]W_{t}[\mathbf{h}]\mathcal{Z%
}[\mathbf{\phi }]\;,  \label{a13}
\end{equation}%
where the weights $W_{s}[\varphi ]$ and $W_{t}[\mathbf{h}]$ are
\begin{eqnarray}
&&W_{s}[\varphi ]=\exp \Big[-\frac{1}{2}\int d\hat{\mathbf{n}}_{1}d\hat{%
\mathbf{n}}_{2}d\mathbf{q}d\tau \varphi _{\mathbf{n}_{1}}^{\ast }(\mathbf{q}%
,\tau )V_{s}^{-1}(\theta _{12})\varphi _{\mathbf{n}_{2}}(\mathbf{q},\tau )%
\Big]\;\;  \label{a13a} \\
&&W_{t}[\mathbf{h}]=\exp \Big[-\frac{1}{2}\int d\hat{\mathbf{n}}_{1}d\hat{%
\mathbf{n}}_{2}d\mathbf{q}d\tau \mathbf{h}_{\mathbf{n}_{1}}^{\dagger }(%
\mathbf{q},\tau )V_{t}^{-1}(\theta _{12})\mathbf{h}_{\mathbf{n}_{2}}(\mathbf{%
q},\tau )\Big]\;\;  \label{a13b}
\end{eqnarray}

Eqs. (\ref{a11}-\ref{a13b}) is the final result of this subsection. We see
that the original problem of the electron with interaction is replaced by a
problem of electron motion in a potential and a magnetic field slowly
varying in space. The condition of the slow variations follows from our
separation into slow pairs. As we will see, at low temperature and weak
interactions the slow variations of the fields $\mathbf{h}$ and $\varphi $
give the main contribution into the physical quantities.

\subsection{Quasiclassical equations}

What we should do is to calculate quantities for any $\mathbf{h}$ and $%
\varphi $ and integrate these quantities over these fields. First, we
introduce the Green functions $G_{\sigma ,\sigma ^{\prime }}^{\phi }\left(
x,x^{\prime }\right) $ corresponding to the action $\mathcal{S}_{eff}\left[
\phi \right] ,$ Eq. (\ref{a12}), as follows

\begin{equation}
G_{\sigma ,\sigma ^{\prime }}^{\phi }\left( x,x^{\prime }\right) =\mathcal{Z}%
^{-1}\left[ \phi \right] \int \chi _{\sigma }\left( x\right) \chi _{\sigma
^{\prime }}^{\ast }\left( x^{\prime }\right) \exp \left( -\mathcal{S}_{eff}%
\left[ \phi \right] \right) D\chi D\chi ^{\ast }
\end{equation}

As the fluctuating field $\phi $, Eq. (\ref{a10a}), varies slowly in space
one can derive quasiclassical equations for the function $G_{\sigma ,\sigma
^{\prime }}^{\phi }\left( x,x^{\prime }\right) $. The method of the
derivation is well known \cite{lo}. One should perform the Fourier transform
with respect to the difference $\mathbf{r-r}^{\prime }$ and assume that the
Green function slowly depends on $\mathbf{R=}\left( \mathbf{r+r}^{\prime
}\right) /2$. Introducing the quasiclassical Green function $g_{\mathbf{n}%
}^{\phi }\left( \mathbf{R,}\tau ,\tau ^{\prime }\right) $ in the standard
way \cite{lo,eilenberger}
\begin{equation}
g_{\mathbf{n}}^{\phi }\left( \mathbf{R,}\tau ,\tau ^{\prime }\right)
=i\int_{-\infty }^{\infty }G_{\mathbf{p}}^{\phi }\left( \mathbf{R,}\tau
,\tau ^{\prime }\right) \frac{d\xi }{\pi },\quad \xi =\frac{\mathbf{p}^{2}}{%
2m}-\epsilon _{F}  \label{a27}
\end{equation}%
we come to the following equation for this function%
\begin{eqnarray}
&&\left( \frac{\partial }{\partial \tau }+\frac{\partial }{\partial \tau
^{\prime }}-iv_{F}\mathbf{n\nabla }\right) g_{\mathbf{n}}^{\phi }\left(
\mathbf{R,}\tau ,\tau ^{\prime }\right)  \nonumber \\
&&+g_{\mathbf{n}}^{\phi }\left( \mathbf{R;}\tau ,\tau ^{\prime }\right) \phi
_{\mathbf{n}}\left( \mathbf{R,}\tau ^{\prime }\right) -\phi _{\mathbf{n}%
}\left( \mathbf{R,}\tau \right) g_{\mathbf{n}}^{\phi }\left( \mathbf{R;}\tau
,\tau ^{\prime }\right) =0  \label{a28}
\end{eqnarray}%
where $\mathbf{n}^{2}=1,$ such that $p_{F}\mathbf{n}$ is a vector on the
Fermi surface.

In principle, one could derive Eq. (\ref{a28}) more accurately, which would
produce additional terms containing space derivatives of the functions $\phi
_{\mathbf{n}}\left( \mathbf{R,}\tau \right) $ and $g_{\mathbf{n}}^{\phi
}\left( \mathbf{R;}\tau ,\tau ^{\prime }\right) $ in the second line.
However, the additional derivatives would compensate infrared singularities
we are interested in. This is the reason why we neglect them. At the same
time, no higher derivatives arise in the first line in Eq. (\ref{a28}) and
this term is exact.

The function $g_{\mathbf{n}}^{\phi }\left( \mathbf{R,}\tau ,\tau ^{\prime
}\right) $ must obey the antiperiodicity conditions
\begin{equation}
g_{\mathbf{n}}^{\phi }\left( \mathbf{R,}\tau ,\tau ^{\prime }\right) =-g_{%
\mathbf{n}}^{\phi }\left( \mathbf{R,}\tau +1/T,\tau ^{\prime }\right) =-g_{%
\mathbf{n}}^{\phi }\left( \mathbf{R,}\tau ,\tau ^{\prime }+1/T\right)
\label{a28a}
\end{equation}%
that follow from Eq. (\ref{a5}).

Eq. (\ref{a28}) is linear and therefore is not sufficient to find $g_{%
\mathbf{n}}^{\phi }\left( \mathbf{R,}\tau ,\tau ^{\prime }\right) $
unambiguously. However, the same equation as Eq. (\ref{a28}) can be written
for the function $g^{2}$
\begin{equation}
g^{2}\left( \mathbf{R,}\tau ,\tau ^{\prime }\right) =\int_{0}^{1/T}g_{%
\mathbf{n}}^{\phi }\left( \mathbf{R,}\tau ,\tau ^{\prime \prime }\right) g_{%
\mathbf{n}}^{\phi }\left( \mathbf{R,}\tau ^{\prime \prime },\tau ^{\prime
}\right) d\tau ^{\prime \prime }  \label{a29}
\end{equation}%
An obvious solution for $g^{2}$ can be written as%
\begin{equation}
g^{2}\left( \mathbf{R,}\tau ,\tau ^{\prime }\right) =c\delta \left( \tau
-\tau ^{\prime }\right)  \label{a30}
\end{equation}%
where $c$ is an arbitrary constant. It can be fixed assuming that the
fermion-fermion interaction is present only in a finite, although
macroscopic part of the space. Then, outside this space we come to the Green
function of a free fermion gas satisfying Eq. (\ref{a30}) with $c=1.$ So, we
come to the equation%
\begin{equation}
\int_{0}^{1/T}g_{\mathbf{n}}^{\phi }\left( \mathbf{R,}\tau ,\tau ^{\prime
\prime }\right) g_{\mathbf{n}}^{\phi }\left( \mathbf{R,}\tau ^{\prime \prime
},\tau ^{\prime }\right) d\tau ^{\prime \prime }=\delta \left( \tau -\tau
^{\prime }\right)  \label{a31}
\end{equation}%
Eq. (\ref{a31}) complements Eq. (\ref{a28}) and these equations are
sufficient to find the function $g_{\mathbf{n}}^{\phi }\left( \mathbf{R,}%
\tau ,\tau ^{\prime }\right) $ for any function $\phi _{\mathbf{p}}\left(
\mathbf{R},\tau \right) $. After that, in order to calculate physical
quantities, one should perform a proper averaging over $\phi _{\mathbf{p}%
}\left( \mathbf{R},\tau \right) $ with the weights $W_{s,t}\left[ \phi %
\right] $, Eqs. (\ref{a13a}, \ref{a13b}). In the next subsection we will
show how to express the partition function $\mathcal{Z}\left[ \phi \right] $%
, Eq. (\ref{a11a}), in terms of the solution of these equations but now let
us reduce Eqs. (\ref{a28}, \ref{a31}) to a more simple form.

Now we come to the main point of the method proposed here. We notice that
the quasiclassical Green function of the free fermion gas is singular and
can be written as
\begin{equation}
g_{0\mathbf{n}}\left( \tau -\tau ^{\prime }\right) =-iTRe\left[ \sin
^{-1}\pi T\left( \tau -\tau ^{\prime }-i\delta \right) \right]  \label{b1}
\end{equation}%
where $\delta \rightarrow +0$. Of course, the Green function $g_{0\mathbf{n}%
}\left( \tau -\tau ^{\prime }\right) $ satisfies Eqs. (\ref{a28}, \ref{a28a}%
, \ref{a31}).

As concerns arbitrary $\phi _{\mathbf{n}}\left( \mathbf{R},\tau \right) ,$
we look for the general solution of Eqs. (\ref{a28}, \ref{a31}) in the
following form

\begin{equation}
g_{\mathbf{n}}^{\phi }\left( \mathbf{R,}\tau ,\tau ^{\prime }\right) =T_{%
\mathbf{n}}\left( \mathbf{R},\tau \right) g_{0}\left( \tau -\tau ^{\prime
}\right) T_{\mathbf{n}}^{-1}\left( \mathbf{R,}\tau ^{\prime }\right)
\label{b3}
\end{equation}%
where $T_{\mathbf{n}}\left( \mathbf{r,}\tau \right) $ is a spin matrix
satisfying the condition%
\begin{equation}
T_{\mathbf{n}}\left( \mathbf{R,}\tau \right) =T_{\mathbf{n}}\left( \mathbf{R,%
}\tau +1/T\right)  \label{b4}
\end{equation}

The representation of the Green function in the form of Eq. (\ref{b3}) is a
generalization of the Schwinger Ansatz \cite{schwinger}. The form given by
Eq. (\ref{b3}), is consistent with Eq. (\ref{a31}) and what remains to be
done is to find the proper matrix $T_{\mathbf{n}}\left( \mathbf{R},\tau
\right) $, such that Eq. (\ref{a28}) is satisfied.

A straightforward manipulation \cite{aleiner} enables one to reduce Eqs. (%
\ref{a28}, \ref{a31}) to the following form%
\begin{equation}
\left( -\frac{\partial }{\partial \tau }+iv_{F}\mathbf{n\nabla }_{\mathbf{R}%
}\right) M_{\mathbf{n}}\left( x\right) +\left[ \phi _{\mathbf{n}}\left(
x\right) ,M_{\mathbf{n}}\left( x\right) \right] =-\frac{\partial \phi _{%
\mathbf{n}}\left( x\right) }{\partial \tau }  \label{b7}
\end{equation}%
where
\begin{equation}
M_{\mathbf{n}}\left( x\right) =\frac{\partial T_{\mathbf{n}}\left( x\right)
}{\partial \tau }T_{\mathbf{n}}^{-1}\left( x\right)  \label{b8}
\end{equation}%
and the symbol $\left[ ,\right] $ stands for the commutator.

Using the representation, Eq. (\ref{a10a}), for the matrix $\phi _{\mathbf{n}%
}\left( x\right) $ and writing the matrix $M_{\mathbf{n}}\left( x\right) $ as%
\begin{equation}
M_{\mathbf{n}}\left( x\right) =\rho _{\mathbf{n}}\left( x\right) +\mathbf{S}%
_{\mathbf{n}}\left( x\right) \mathbf{\sigma ,}  \label{b9}
\end{equation}%
where $\rho _{\mathbf{n}}\left( x\right) $ is a scalar function and $\mathbf{%
S}_{\mathbf{n}}\left( x\right) $ is a three dimensional vector, we reduce
Eq. (\ref{b7}) to two independent equations for $\rho _{\mathbf{n}}\left(
x\right) $ and $\mathbf{S}_{\mathbf{n}}\left( x\right) $%
\begin{equation}
\left( -\frac{\partial }{\partial \tau }+iv_{F}\mathbf{n\nabla }_{\mathbf{R}%
}\right) \rho _{\mathbf{n}}\left( x\right) =-i\frac{\partial \varphi _{%
\mathbf{n}}\left( x\right) }{\partial \tau }  \label{b10}
\end{equation}%
\begin{equation}
\left( -\frac{\partial }{\partial \tau }+iv_{F}\mathbf{n\nabla }_{\mathbf{R}%
}\right) \mathbf{S}_{\mathbf{n}}\left( x\right) +2i\left[ \mathbf{h}_{%
\mathbf{n}}\left( x\right) \mathbf{\times S}_{\mathbf{n}}\left( x\right) %
\right] =-\frac{\partial \mathbf{h}_{\mathbf{n}}\left( x\right) }{\partial
\tau }  \label{b11}
\end{equation}%
Eqs. (\ref{b10}, \ref{b11}) are the final quasiclassical equations that will
be used for further calculations. We emphasize that Eqs. (\ref{b10}, \ref%
{b11}) are obtained from Eqs. (\ref{a28}, \ref{a31}) without making any
approximation. The variable $\rho _{\mathbf{n}}\left( x\right) $ describes
collective density excitations, whereas the variable $\mathbf{S}_{\mathbf{n}%
}\left( x\right) $ stands for spin ones.

Eqs. (\ref{b10}, \ref{b11}) describing these excitations are remarkably
different from each other. Eq. (\ref{b10}) for the density is rather simple.
This is what one obtains using the high dimensional bosonization of Refs.
\cite%
{haldane,houghton1,houghton,neto,kopietz,kopietzb,khveshchenko,khveshchenko1,metzner}
from an eikonal equation. Of course, we could take into account gradients of
the field $\varphi _{\mathbf{n}}\left( x\right) $ and this would lead to
additional terms in the L.H.S. of Eq. (\ref{b10}). However, this does not
lead to new physical effects.

In contrast, Eq. (\ref{b11}) is highly non-trivial due to the presence of $%
\mathbf{h}_{\mathbf{n}}\left( x\right) $ in the L.H.S of the equation.
Actually, the homogeneous part of Eq. (\ref{b11}) is just the equation of
motion of a classical magnetic moment in the external magnetic field. We
will see that the form of Eq. (\ref{b11}) will result in very non-trivial
effects that will be considered later.

The presence of the second term in the L.H.S. of Eq. (\ref{b11}) is a
consequence of a non-abelian character of the variables describing the spin
excitations. In this respect our method resembles the non-abelian
bosonization well known for one-dimensional systems \cite{tsvelik}.

It is important to emphasize that the variables $\rho _{\mathbf{n}}\left(
x\right) $ and \ $\mathbf{S}_{\mathbf{n}}\left( x\right) $ depend not only
on the time and coordinate but also on the position of the vector $\mathbf{n}
$, that determines the position on the Fermi surface. This dependence is
usual for kinetic equations. In the previous attempts to construct a higher
dimensional bosonization \cite%
{haldane,houghton1,houghton,neto,kopietz,kopietzb,khveshchenko,khveshchenko1,metzner}
the corresponding variable arose from a \textquotedblleft
patching\textquotedblright\ of the Fermi surface.

\subsection{Partition function.}

What remains to be done is to express the partition function $\mathcal{Z}%
\left[ \Phi \right] $, Eq. (\ref{a11a}, \ref{a12}), in terms of the
variables $\rho _{\mathbf{n}}\left( x\right) $ and $\mathbf{h}_{\mathbf{n}%
}\left( x\right) $. Integrating over $\chi $, $\chi ^{\ast }$ in Eq. (\ref%
{a11a}) and using Eqs. (\ref{a12}, \ref{a3}) for $\mathcal{S}_{eff}\left[
\phi \right] $ we write $\mathcal{Z}\left[ \phi \right] $ in the form%
\begin{equation}
\mathcal{Z}\left[ \phi \right] =\exp \left[ Tr\int \ln \left( -\frac{%
\partial }{\partial \tau }-\frac{\mathbf{\hat{p}}^{2}}{2m}+\epsilon
_{F}+\phi _{\mathbf{n}}\left( x\right) \right) dx\right]  \label{b12}
\end{equation}%
Eq. (\ref{b12}) can further be represented as%
\begin{equation}
\mathcal{Z}\left[ \phi \right] =Z_{0}\exp \left[ Tr\int \int_{0}^{1}G^{\phi
}\left( x,x;u\right) \phi _{\mathbf{n}}\left( x\right) dxdu\right]
\label{b13}
\end{equation}%
where the Green function $G^{\phi }\left( \mathbf{r,r;}\tau ,\tau ;u\right) $
is the solution of the equation
\begin{equation}
\left( -\frac{\partial }{\partial \tau }-\frac{\mathbf{\hat{p}}^{2}}{2m}%
+\epsilon _{F}+u\phi _{\mathbf{n}}\left( x\right) \right) G^{\phi }\left(
x,x^{\prime };u\right) =\delta \left( x-x^{\prime }\right)  \label{b14}
\end{equation}%
We see that the calculation of the partition function reduces to the
calculation of the Green function $G^{\phi }\left( x,x^{\prime };u\right) $
that differs from the Green functions calculated in the preceding subsection
by the replacement $\phi \left( x\right) \rightarrow u\phi \left( x\right) $%
, where $u$ is a parameter in the interval $\left( 0,1\right) $. This means
that calculating the Green function we can repeat all the transformations we
have performed previously.

The Green function $G^{\phi }\left( \mathbf{r},\tau ;\mathbf{r},\tau
^{\prime };u\right) $ at coinciding points can be written in terms of the
quasiclassical Green function \ $g_{\mathbf{n}}^{\phi }$ as
\begin{equation}
G^{\phi }\left( x,x^{\prime };u\right) =-i\pi \nu \int g_{\mathbf{n}}^{\phi
}\left( \mathbf{r;}\tau ,\tau ^{\prime }\mathbf{;}u\right) d\mathbf{n}
\label{b16}
\end{equation}%
where $\nu $ is the density of states on the Fermi surface (without taking
into account the double degeneracy due to spin).

Next, we use the representation, Eq. (\ref{b3}), and expand the function $T_{%
\mathbf{n}}^{-1}\left( \mathbf{r},\tau ^{\prime }\right) $ in $\tau ^{\prime
}-\tau $. The contribution from $g_{0\mathbf{n}}\left( \tau -\tau ^{\prime
}\right) $, Eq. (\ref{b1}), vanishes at $\tau =\tau ^{\prime }$. Using Eqs. (%
\ref{b8}, \ref{b9}) we obtain finally%
\begin{equation}
\mathcal{Z}\left[ \phi \right] =\mathcal{Z}_{0}\mathcal{Z}_{\varphi }%
\mathcal{Z}_{\mathbf{h}}  \label{b17}
\end{equation}%
where%
\begin{equation}
\mathcal{Z}_{\varphi }=\exp \left[ -2i\nu \int \int_{0}^{1}\rho _{\mathbf{n}%
}\left( x,u\right) \varphi \left( x\right) dxd\mathbf{n}du\right] ,
\label{b18}
\end{equation}%
\begin{equation}
\mathcal{Z}_{\mathbf{h}}=\exp \left[ -2\nu \int \int_{0}^{1}\mathbf{S}_{%
\mathbf{n}}\left( x,u\right) \mathbf{h}_{\mathbf{n}}\left( x\right) dxd%
\mathbf{n}du\right]  \label{b19}
\end{equation}%
The functions $\rho _{\mathbf{n}}\left( x,u\right) $ and $\mathbf{S}_{%
\mathbf{n}}\left( x,u\right) $ should be found from the equations

\begin{equation}
\hat{L}_{u=0}\left( \mathbf{n}\right) \rho _{\mathbf{n}}\left( x\right) =-iu%
\frac{\partial \varphi _{\mathbf{n}}\left( x\right) }{\partial \tau },\quad
\hat{L}_{u}\left( \mathbf{n}\right) \mathbf{S}_{\mathbf{n}}\left( x\right)
=-u\frac{\partial \mathbf{h}_{\mathbf{n}}\left( x\right) }{\partial \tau }
\label{b20}
\end{equation}%
In Eqs. (\ref{b20}), the operator $\hat{L}_{u}$ has the form

\begin{equation}
\hat{L}_{u}\left( \mathbf{n}\right) =-\frac{\partial }{\partial \tau }%
+iv_{F}\left( \mathbf{n\nabla }_{\mathbf{r}}\right) +2iu\hat{h}  \label{b23}
\end{equation}%
where the matrix $\hat{h}$ is%
\begin{equation}
\hat{h}=\left(
\begin{array}{ccc}
0 & -h_{z} & h_{y} \\
h_{z} & 0 & -h_{x} \\
-h_{y} & h_{x} & 0%
\end{array}%
\right)  \label{b24}
\end{equation}%
and $h_{x},h_{y},$ and $h_{z}$ are the components of the vector $\mathbf{h}$
($\hat{h}\mathbf{a=}\left[ \mathbf{h\times a}\right] $ for any vector $%
\mathbf{a}$).

The functions $\mathbf{S}_{\mathbf{n}}\left( x\right) $ and $\rho _{\mathbf{n%
}}\left( x\right) $ are periodic in $\tau$%
\begin{equation}
\mathbf{S}_{\mathbf{n}}\left( \mathbf{r,}\tau\right) =\mathbf{S}_{\mathbf{n}%
}\left( \mathbf{r,}\tau+1/T\right)  \label{b24a}
\end{equation}

The accuracy of Eqs. (\ref{a13a}, \ref{a13b}) can be somewhat improved by
making the substitution \ $V_{s,t}\rightarrow \hat{\Gamma}_{s,t}/\nu $,
where \ $\hat{\Gamma}_{s,t}$ is the scattering amplitude for the singlet and
triplet channel respectively.

Thus, we have reduced the study of the system of the interacting fermions to
investigation of a system of bosonic density and spin excitations. Therefore
the word \textquotedblleft bosonization\textquotedblright\ is most suitable
for our approach. We see that the method should work in any dimension. At
the same time, it is more general than the scheme of the high dimensional
bosonization of Refs. \cite%
{haldane,houghton1,houghton,neto,kopietz,kopietzb,khveshchenko,khveshchenko1,metzner}
because we can consider the spin excitations that are much more interesting
than the density ones. The presence of the non-trivial third term in Eqs. (%
\ref{b11}, \ref{b23}) is a consequence of the non-abelian character of the
excitations. In contrast, the previous schemes worked only for abelian
density excitations and it is simply impossible to write any interaction of $%
\rho $ with an external field $\varphi $ without using time or space
derivatives.

\section{Effective field theory and renormalization group equations}

\label{field theory}

\subsection{Infrared logarithmic divergences.}

Eqs. (\ref{b17}-\ref{b24a}), (\ref{a13}-\ref{a13b}) are sufficient for
calculating low energy contributions to the thermodynamic quantities.
Calculation for the density excitations is not difficult because the first
equation (\ref{b20}) can easily be solved explicitly. As concerns the spin
excitations one may seek for the solution of the second equation (\ref{b20})
expanding the operator $L_{u}$, Eq. (\ref{b20}),\ in $\hat{h}$. It turns out
that in the limit $T\rightarrow 0$\ terms of the perturbation theory for
scattering amplitudes are logarithmically divergent in any dimension and one
has to sum an infinite series.

In order to see the origin of this divergency, let us consider the
expression $K\left( \mathbf{n,-n}^{\prime }\right) =\int L_{u}^{-1}\left(
\mathbf{n;}x,x^{\prime }\right) L_{u}^{-1}\left( -\mathbf{n}^{\prime }%
\mathbf{;}x^{\prime },x\right) dx^{\prime }$.\ Using the Fourier transform
in the coordinates and time we bring this expression in the limit $%
T\rightarrow 0$ to the form%
\begin{equation}
K\left( \mathbf{n,-n}^{\prime }\right) =\int \frac{d\omega d^{d}\mathbf{k}}{%
\left( 2\pi \right) ^{d+1}}\frac{1}{i\omega -v_{F}\mathbf{kn}}\frac{1}{%
i\omega +v_{F}\mathbf{kn}^{\prime }}  \label{b100}
\end{equation}%
If $\mathbf{n}$ is parallel to $\mathbf{n}^{\prime }$ we can integrate
separately over the parallel $k_{\parallel }$ and perpendicular $\mathbf{k}%
_{\perp }$ (with respect to the vector $\mathbf{n}$) components of the
vector \ $\mathbf{k}$. In this case the integrand does not depend on $%
\mathbf{k}_{\perp }$ and formally diverges at large $\left\vert \mathbf{k}%
_{\perp }\right\vert $. However, we assumed that the momenta cannot be very
large and therefore the maximal $\left\vert \mathbf{k}_{\perp }\right\vert $
are of the order of $r_{0}^{-1}$ from Eq. (\ref{a7c}), which provides the
convergence of the integral over $\mathbf{k}_{\perp }$.

In contrast, the integral over $\omega $ and $k_{\parallel }$ diverges at $%
\emph{small}$ values of these variables. These are infrared divergences and
they lead to important contributions to the thermodynamic quantities.
Estimating the value of \ $K$ we come to the following expression
\begin{equation}
K\left( \mathbf{n,-n}^{\prime }\right) \sim r_{0}^{1-d}v_{F}^{-1}\ln \left(
\max \left\{ Tv_{F}^{-1}r_{0},\left\vert \mathbf{n-n}^{\prime }\right\vert
\right\} \right)  \label{b101}
\end{equation}%
Products of the function $K\left( \mathbf{n,-n}^{\prime }\right) $ arise
when calculating scattering amplitudes with the help of the perturbation
theory and we see that in the limit $T\rightarrow 0$ one should sum an
infinite series. This is not an easy task but the consideration simplifies
if we reformulate the problem of calculation of the partition function in
terms of a field theory. Then we will be able to use for summation of the
logarithms a renormalization group technique.

\subsection{Low energy supersymmetric field theory.}

Now, our task is to solve Eqs. (\ref{b20}), substitute the solution into
Eqs. (\ref{b17}, \ref{b19}) and, using the obtained expression for $\mathcal{%
Z}\left[ \phi \right] $, calculate the integral over the fields $\varphi _{%
\mathbf{n}}$ and $\mathbf{h}_{\mathbf{n}}$. This procedure is somewhat
similar to what one does in theory of disordered metals. A convenient way of
calculations is to represent the solutions of Eqs. (\ref{b20}) in a form
that would allow one to integrate over the fields $\varphi _{\mathbf{n}}$ in
the beginning of all the calculations. Integration over supervectors, Refs.
\cite{review,book}, is most convenient for this purpose and we follow now
this method. As the contribution of the density fluctuations is simple, we
consider from now on the spin excitations only. The contribution of the
density excitations will be added in final results.

Using supervectors $\psi $ and formulae for gaussian integration we
represent the partition function $\mathcal{Z}_{\mathbf{h}},$ Eq. (\ref{b19}%
), as follows \cite{aleiner}%
\begin{equation}
\mathcal{Z}_{\mathbf{h}}=\int \exp [2\nu \sqrt{2i}\int \overline{\psi \left(
X\right) }\mathcal{F}\left( X\right) dX]\exp \left[ -\mathcal{S}_{\mathbf{h}}%
\left[ \psi \right] \right] D\psi  \label{c33}
\end{equation}%
where

\begin{equation}
\mathcal{S}_{\mathbf{h}}\left[ \psi \right] =-2i\nu \int \overline{\psi
\left( X\right) }\mathcal{H}\psi \left( X\right) dX,\quad \ \mathcal{H}=%
\mathcal{H}_{0}-2iu\hat{h}\left( X\right) \tau _{3}  \label{c1}
\end{equation}%
The weight denominator in the gaussian integral, Eq. (\ref{c33}), is absent
because $\psi $ are supervectors. Although the general form of Eqs. (\ref%
{c33}, \ref{c1}) is simple (it is a gaussian integral), the detailed
structure of the vectors and matrices is not. The supervectors $\psi $ have $%
48$ components. The number of components comes from the necessity to
consider 1) $3$ spin components ($s$-space), 2) bosonic and fermionic
variables ($g$-space), 3) \textquotedblleft particles\textquotedblright\ and
\textquotedblleft holes\textquotedblright\ ($eh$-space) , 4)
\textquotedblleft left\textquotedblright\ and \textquotedblleft
right\textquotedblright\ motion ($n$-space), 5) one should double the number
of the components to \textquotedblleft hermitize\textquotedblright\ the
space ($H$-space). The operator $\hat{L}_{u}\left( \mathbf{n}\right) $, Eq. (%
\ref{b23}), is not hermitian and the \textquotedblleft
hermitization\textquotedblright\ of the space of the supervectors is
necessary to provide convergence of the gaussian integral in Eq. (\ref{c33}).

The supervectors $\psi $ are assumed to satisfy the bosonic periodicity
conditions%
\begin{equation}
\psi \left( \tau \right) =\psi \left( \tau +1/T\right)  \label{c1a}
\end{equation}%
We emphasize that Eq. (\ref{c1a}) holds for both the bosonic and fermionic
components of the supervectors $\psi .$

The generalized coordinate $X$ contains the components

\begin{equation}
X=\left( x,z\right) ,\quad z=\left( \mathbf{n,}u\right)  \label{c2}
\end{equation}%
where $\mathbf{n}$ is now a unit vector, $\mathbf{n}^{2}=1$, parallel to the
original vector $\mathbf{n}$ but having only positive $x$ components $%
n_{x}>0 $. Negative $x$-components are taken into account by doubling the
number the components of the supervector $\psi $ ($n$-space).

The operator $\mathcal{H}_{0}$ in Eq. (\ref{c1}) can be written as
\begin{equation}
\mathcal{H}_{0}=-iv_{F}\left( \mathbf{n\nabla }\right) \tau _{3}\Sigma
_{3}-\Lambda _{1}\frac{\partial }{\partial \tau }  \label{c29}
\end{equation}%
where $\Lambda _{1}$ is the first Pauli matrix in the hermitized space (it
is unity in all the other spaces) and the matrices $\tau _{3}$ and $\Sigma
_{3}$ are the third Pauli matrices in the $eh$ and $n$ spaces respectively%
\begin{equation}
\Lambda _{1}=\left(
\begin{array}{cc}
0 & 1 \\
1 & 0%
\end{array}%
\right) _{H},\quad \tau _{3}=\left(
\begin{array}{cc}
1 & 0 \\
0 & -1%
\end{array}%
\right) _{eh},\quad \Sigma _{3}=\left(
\begin{array}{cc}
1 & 0 \\
0 & -1%
\end{array}%
\right) _{n}  \label{c30}
\end{equation}

The matrix $\hat{h}\left( X\right) $ has the spin structure of Eq. (\ref{b24}%
) (the replacement $\mathbf{n\rightarrow n}\Sigma _{3}$ should be made) and
is unity in all other spaces. The action $\mathcal{S}_{\mathbf{h}}\left[
\psi \right] $\textrm{\ }is supersymmetric and is invariant under
homogeneous rotations in the superspace.

The supervector $\bar{\psi}$ in Eq. (\ref{c33}) is conjugated with respect
to $\psi $ (see for the definition Ref. \cite{aleiner}). The vector $%
\mathcal{F}$ has the form%
\begin{equation}
\mathcal{F}_{\mathbf{h}}\left( X\right) =\mathcal{F}_{\mathbf{h}}^{1}\left(
X\right) +\mathcal{F}_{\mathbf{h}}^{2}\left( X\right)  \label{c5}
\end{equation}%
where%
\[
\mathcal{F}_{\mathbf{h}\gamma }^{1}=\frac{1}{\sqrt{2}}\left(
\begin{array}{c}
0 \\
1%
\end{array}%
\right) _{g}\otimes \left(
\begin{array}{c}
1 \\
1%
\end{array}%
\right) _{H}\otimes \left(
\begin{array}{c}
0 \\
1%
\end{array}%
\right) _{eh}\otimes \left(
\begin{array}{c}
h_{\gamma }\left( \mathbf{n}\right) \\
h_{\gamma }\left( -\mathbf{n}\right)%
\end{array}%
\right) _{n}
\]%
\[
\mathcal{F}_{\mathbf{h}\gamma }^{2}=\frac{1}{\sqrt{2}}\left(
\begin{array}{c}
0 \\
1%
\end{array}%
\right) _{g}\otimes \left(
\begin{array}{c}
1 \\
-1%
\end{array}%
\right) _{H}\otimes \left(
\begin{array}{c}
1 \\
0%
\end{array}%
\right) _{eh}\otimes \left(
\begin{array}{c}
u\partial _{\tau }h_{\gamma }\left( \mathbf{n}\right) \\
u\partial _{\tau }h_{\gamma }\left( -\mathbf{n}\right)%
\end{array}%
\right) _{n}
\]%
and $\gamma =x,y,z$ stands for the spin indices.

We see (\ref{c33}, \ref{c1}, \ref{c5}) that the Hubbard-Stratonovich field $%
h\left( X\right) $ enters both $\mathcal{H}$ and $\mathcal{F}$ linearly. We
write the contribution of the spin excitations $\mathcal{Z}_{t}$ into the
partition function as%
\begin{equation}
\mathcal{Z}_{t}=\int \mathcal{Z}_{\mathbf{h}}W_{t}\left[ \mathbf{h}\right]
\mathcal{D}\mathbf{h,}  \label{c6}
\end{equation}%
(c.f. Eq. (\ref{a13})), which enables us to integrate immediately over $%
\mathbf{h}$.

The integration over $\mathbf{h}$ is gaussian and can easily be performed.
As a result, we obtain an effective action $\mathcal{S}$ containing non only
the free quadratic part but also cubic and quartic interactions
\begin{equation}
\mathcal{Z}_{t}=\int \mathcal{D}{\psi }\;\exp \left( -\mathcal{S}\right)
,\quad \mathcal{S}=\mathcal{S}_{0}+\sum_{a=1,2,3}\mathcal{S}_{a}  \label{c7}
\end{equation}%
The interaction-independent part $\mathcal{S}_{0}$ equals
\begin{equation}
\mathcal{S}_{0}=-2i\nu \int dX\;\overline{{\psi }_{\alpha }}\left( X\right)
\mathcal{H}_{0}{\psi }_{\alpha }\left( X\right)  \label{c8}
\end{equation}%
The three different interaction terms present in the theory can be written
as
\begin{eqnarray}
\mathcal{S}_{2} &=&-i\nu \sum_{i,j=1}^{4}\lambda _{ij}\int dXdX_{1}
\label{c9} \\
&&\times \left( \overline{\psi }_{\delta }\left( X\right) \tau _{3}\Pi
_{j}\partial _{X}\mathcal{F}_{0}\right) \;{\Gamma }_{X,X_{1}}^{i}\;\left(
\overline{\mathcal{F}_{0}}\overline{\partial _{X_{1}}}\Pi _{j}\tau _{3}\psi
_{\delta }\left( X_{1}\right) \right)  \nonumber
\end{eqnarray}%
\begin{eqnarray}
\mathcal{S}_{3} &=&-2\sqrt{-2i}\nu \sum_{i,j=1}^{4}\lambda
_{ij}\;\varepsilon _{\delta \beta \gamma }  \label{c10} \\
&&\times \int dXdX_{1}\left( \overline{\psi }_{\delta }\left( X\right) u\tau
_{3}\Pi _{j}\psi _{\beta }\left( X\right) \right) \;{\Gamma }%
_{X,X_{1}}^{i}\left( \overline{\mathcal{F}_{0}}\overline{\partial _{X_{1}}}%
\tau _{3}\psi _{\gamma }\left( X_{1}\right) \right)  \nonumber
\end{eqnarray}%
\begin{eqnarray}
\mathcal{S}_{4} &=&-2\nu \sum_{i,j=1}^{4}\lambda _{ij}\;\varepsilon _{\delta
\beta \gamma }\varepsilon _{\delta \beta _{1}\gamma _{1}}  \label{c11} \\
&&\times \int dXdX_{1}\left( \overline{\psi }_{\beta }\left( X\right) u\tau
_{3}\Pi _{j}\psi _{\gamma }\left( X\right) \right) {\Gamma }%
_{X,X_{1}}^{i}\left( \overline{\psi }_{\beta _{1}}\left( X\right) u_{1}\tau
_{3}\Pi _{j}\psi _{\gamma _{1}}\left( X_{1}\right) \right)  \nonumber
\end{eqnarray}%
Summation over spin indices $\alpha ,\beta ,\delta ,\gamma $ is implied and
we use the absolutely antisymmetric tensor $\varepsilon _{\alpha \beta
\gamma }$ with $\varepsilon _{123}=1$. The action $\mathcal{S}$, Eqs. (\ref%
{c7}-\ref{c11}), is sufficient for calculation of the thermodynamic
potential in the absence of a magnetic field. The operator ${\Gamma }%
_{X,X^{\prime }}^{i}$ has the form
\begin{equation}
{\Gamma }_{X,X^{\prime }}^{i}=\gamma _{i}\left( \widehat{\mathbf{n}\mathbf{n}%
^{\prime }}\right) \;f(\mathbf{r}-\mathbf{r}^{\prime })\;\delta (\tau -\tau
^{\prime })  \label{c12}
\end{equation}%
and
\begin{eqnarray}
\gamma _{1}(\widehat{\mathbf{n}\mathbf{n}_{1}}) &=&\left( \frac{\nu \hat{V}%
_{t}}{1-2\nu \hat{V}_{t}}\right) (\widehat{\mathbf{n},\mathbf{n}_{1}})\equiv
\gamma _{f}  \label{eq:4.39} \\
\gamma _{2}(\widehat{\mathbf{n}\mathbf{n}_{1}}) &=&\left( \frac{\nu \hat{V}%
_{t}}{1-2\nu \hat{V}_{t}}\right) (\widehat{\mathbf{n},-\mathbf{n}_{1}}%
)\equiv \gamma _{b}  \label{eq:4.40}
\end{eqnarray}%
where $f\left( \mathbf{r}\right) $ is the cutoff function introduced in Eqs.
(\ref{a7a}, \ref{a7c}). The operator $\partial _{X}$ in Eqs. (\ref{c9}-\ref%
{c11}) has the form
\begin{equation}
\partial _{X}(\alpha )=\left(
\begin{array}{cc}
1 & 0 \\
0 & u\partial _{\tau }%
\end{array}%
\right) _{eh}  \label{c12a}
\end{equation}%
As we will see, the most interesting contribution comes from $\mathbf{n}$
and $\mathbf{n}_{1}$ nearly parallel to each other. This justifies the
notations $\gamma _{f}$ and $\gamma _{b}$ standing for the bare forward and
backward scattering.

The matrices $\Pi _{i}$ equal%
\begin{equation}
\Pi _{1}=1,\quad \Pi _{2}=\Sigma _{3},\quad \Pi _{3}=\Lambda _{1}\tau
_{3},\quad \Pi _{4}=\Lambda _{1}\tau _{3}\Sigma _{3}  \label{c13}
\end{equation}%
and
\begin{equation}
\lambda _{ij}=\left(
\begin{array}{cccc}
1 & 1 & -1 & -1 \\
1 & 1 & 1 & 1 \\
1 & -1 & 1 & -1 \\
1 & -1 & -1 & 1%
\end{array}%
\right)  \label{c14}
\end{equation}

The vector $\mathcal{F}_{0}$ in Eqs. (\ref{c9}-\ref{c11}) has the form%
\begin{equation}
\mathcal{F}_{0}=\frac{1}{\sqrt{2}}\left(
\begin{array}{c}
0 \\
1%
\end{array}%
\right) _{g}\left(
\begin{array}{c}
1 \\
1%
\end{array}%
\right) _{n}\left(
\begin{array}{c}
\left(
\begin{array}{c}
1 \\
1%
\end{array}%
\right) _{eh} \\
\left(
\begin{array}{c}
-1 \\
1%
\end{array}%
\right) _{eh}%
\end{array}%
\right) _{H}  \label{c15}
\end{equation}%
The vector $\mathcal{F}_{0}$ projects on the bosonic sector and its presence
violates the supersymmetry of the terms $\mathcal{S}_{a}$, $a=2,3$.

The action $\mathcal{S}$, (\ref{c8}-\ref{c11}) has the cubic and quartic
interaction terms and looks quite complicated. Nevertheless, explicit
calculations are not very difficult because perturbation theory in the
interaction is logarithmic (see Eqs. (\ref{b100}, \ref{b101}) in all
dimensions $d=1,2,3$. In order to sum up the logarithms we use a
renormalization group scheme.

At the end of this subsection we write additional terms in the action
arising due to an external magnetic field $\mathbf{b}$ acting on spins.
Writing these terms is necessary for calculation of the spin susceptibility.
These terms can be brought to the form \cite{schwiete}

\begin{eqnarray}
&&\mathcal{S}_{b0}=-\nu \;\eta \;\int dx\;\mathbf{b}^{2}(x)  \label{c16} \\
&&\mathcal{S}_{b1}=-2\nu \sqrt{-2i}\;\eta \;\int dX\;b_{\delta }(x)\left(
\overline{\psi }_{X,\delta }\tau _{3}\partial _{X}\mathcal{F}_{0}\right)
\qquad  \label{eq:4.46} \\
&&\mathcal{S}_{b2}=4\nu \;\varepsilon _{\delta \beta \gamma }\;\eta \;\int
dX\;b_{\delta }(x)\left( \overline{\psi }_{X,\beta }u\tau _{3}\psi
_{X,\gamma }\right)  \label{c17}
\end{eqnarray}%
In these expressions
\begin{equation}
\eta =\frac{1}{1-2\nu \overline{V_{t}}},  \label{eq:4.48}
\end{equation}%
where the bar in $\overline{V_{t}}$ means averaging over the full solid
angle.

\subsection{Renormalization group equations and their solutions.}

We use a standard momentum shell renormalization group scheme. Separating
fast and slow fields in the action we integrate over the fast fields and
determine in this way the flow of coupling constants as a function of a
running cutoff. In our case this amounts to a re-summation of the
perturbation theory in the leading logarithmic approximation. We assume
during the renormalization that the coupling constants $\gamma $ are small, $%
\gamma \ll 1$.

In our case it convenient to define fast ${\Phi }$ and slow $\Psi $ fields
with respect to the frequency only. The reason is the anisotropy in
momentum. As one can see, relevant momenta $p_{\parallel }$ are of the order
of $\omega /v_{F}$, while momenta $\mathbf{p}_{\perp }$ do not contribute to
the logarithm and enter as parameters. Thus, we write
\begin{equation}
{\psi }(X)=\Psi (X)+{\Phi }(X),  \label{c40}
\end{equation}%
where the fast fields ${\Phi }$ have the frequencies $\omega $ in the
interval,
\[
\kappa \omega _{c}<|\omega |<\omega _{c}
\]%
while the slow ones $\Psi $ carry frequencies
\[
|\omega |<\kappa \omega _{c},
\]%
where $\omega _{c}$ is the running cut-off and $\kappa <1$. Fast modes are
integrated over in the Gaussian approximation using averages of the form
\begin{equation}
\left\langle \dots \right\rangle _{0}=\int d{\Phi }\;(\dots )\exp \left( -%
\mathcal{S}_{0}[{\Phi }]\right) .  \label{c41}
\end{equation}%
This results in a change in $\mathcal{S}$
\begin{equation}
\delta \mathcal{S}[\Psi ]=-\ln \left\langle \exp \left( -\mathcal{S}[\Psi +{%
\Phi }]\right) \right\rangle _{0}-\mathcal{S}[\Psi ],  \label{c42}
\end{equation}%
that will now be determined explicitly.

The main object of the RG calculations that we start now is the Green
function $\mathcal{G}_{0}\left( \mathbf{k,n,}\omega \right) $ corresponding
to the bare action $\mathcal{S}_{0}$, Eq. (\ref{c8}). Calculating a gaussian
integral we obtain easily the form of the Green function in the Fourier
representation
\begin{equation}
\mathcal{G}_{0}\left( \mathbf{k,n,}\omega \right) =-4i\nu \left\langle \psi
\bar{\psi}\right\rangle _{0}=\frac{1}{i\omega \Lambda _{1}+v_{F}\mathbf{kn}%
\tau _{3}\Sigma _{3}}  \label{d1}
\end{equation}%
where $\left\langle ...\right\rangle _{0}$ stands for averaging with $%
\mathcal{S}_{0},$ Eq. (\ref{c8}).

The bare Green function, Eq. (\ref{d1}) has a non-trivial matrix structure.
Therefore renormalized vertices are also complicated and the form of the
quadratic $\mathcal{S}_{2}$ and cubic terms $\mathcal{S}_{3}$ in the action,
Eqs. (\ref{c9}-\ref{c10}), is not most general. At the same time, the
supersymmetric terms $\mathcal{S}_{0}$ and $\mathcal{S}_{4}$ do not change
their form under the renormalization. Actually, $\mathcal{S}_{0}$ does not
change in the first order and therefore we write the corresponding equations
first for the effective vertices $\gamma _{i}\left( \xi \right) ,$ $%
i=1,2,3,4,$ of the quartic interaction, where $\xi $ is the running
logarithmic variable.

A detailed calculation has been performed in Ref. \cite{aleiner}. It turns
out that only the vertices $\gamma _{1}$ and $\gamma _{3}$ are non-trivially
renormalized, whereas $\gamma _{2}$ and $\gamma _{4}$ remain equal to their
bare values $\gamma _{2}^{0}=\gamma _{f}$ and $\gamma _{4}^{0}=\gamma _{b}$.
At the same time, only $\gamma _{3}\left( \xi \right) $ enters the
thermodynamic potential and we write here the RG equation for $\gamma
_{3}\left( \xi \right) $ only%
\begin{equation}
\frac{d{\gamma }_{3}\left( \xi \right) }{d\xi }=-\left[ {\gamma }_{3}(\xi )%
\right] ^{2}  \label{d3}
\end{equation}

The renormalization of the $\mathcal{S}_{2}$ and $\mathcal{S}_{3}$ terms is
more complicated. In order to obtain a closed system of RG equations one
should write a more general form of the quadratic and cubic interactions.
This is because the renormalization of different elements in the
particle-hole space runs in a different way. As a result \cite{aleiner},
there are $2$ different cubic vertices $\beta _{i}^{+}$ and $\beta _{i}^{-}$
and $3$ quadratic vertices $\Delta _{i}^{++}$, $\Delta _{i}^{--}$ and $%
\Delta _{i}^{+-}=\Delta _{i}^{-+}$ for all $i=1,2,3,4$. The bare values of
all these vertices can be found from Eqs. (\ref{c9}, \ref{c10}) and they are
equal to $\gamma _{f}$ for $i=1,2$ and $\gamma _{b}$ for $i=3,4$.

The vertices $\beta _{i}$ and $\Delta _{i}$ do not renormalize for $i=2,4$
and remain equal to their bare values. Moreover, the vertices $\Delta _{i}$
do not renormalize for $i=1,$ too. The thermodynamic potential is determined
solely by $\Delta _{3}\left( \xi \right) $ and we write down the RG
equations for $i=3$ only. We remind the reader that this value of $i$
corresponds to the backward scattering.

The equations for the cubic vertices $\beta _{3}^{\pm }\left( \xi \right) $
can be written as
\begin{equation}
\frac{d{\beta }_{3}^{+}\left( \xi \right) }{d\xi }=-2{\gamma }_{3}\left( \xi
\right) {\beta }_{3}^{+}\left( \xi \right) ;\;\frac{d{\beta }_{3}^{-}\left(
\xi \right) }{d\xi }=-{\gamma }_{3}\left( \xi \right) {\beta }_{3}^{-}\left(
\xi \right) ;  \label{d4}
\end{equation}%
The equation for $\Delta _{3}^{++}\left( \xi \right) $ takes the form%
\begin{equation}
\frac{d\Delta _{3}^{++}\left( \xi \right) }{d\xi }=-2\Delta _{3}^{++}\left(
\xi \right) \gamma _{3}\left( \xi \right) -2\left[ \beta _{3}^{+}\left( \xi
\right) \right] ^{2};  \label{d5}
\end{equation}%
whereas the equation for $\Delta _{3}^{\pm }\left( \xi \right) $ is
\begin{equation}
\frac{d\Delta _{3}^{-+}\left( \xi \right) }{d\xi }=\frac{d\Delta
_{3}^{+-}\left( \xi \right) }{d\xi }=-2\beta _{3}^{-}\left( \xi \right)
\beta _{3}^{+}\left( \xi \right) .  \label{d6}
\end{equation}%
Equation for $\Delta _{3}^{--}\left( \xi \right) $ can be written from the
condition of the absence of ultraviolet divergencies, which follows from the
invariance of the system under spin rotations. It takes the form
\begin{equation}
\Delta _{3}^{--}\left( \xi \right) \gamma _{3}(\xi )=\left[ \beta
_{3}^{-}\left( \xi \right) \right] ^{2}  \label{d7}
\end{equation}

Eqs. (\ref{d3}-\ref{d7}) can easily be solved and their solutions satisfying
the boundary conditions at $\xi =0$ (when the vertices are equal to their
bare values) take the form
\begin{eqnarray}
&&\gamma _{3}\left( \xi \right) =\beta _{3}^{-}\left( \xi \right) =\Delta
_{3}^{--}\left( \xi \right) =\frac{1}{\xi _{b}^{\ast }+\xi };  \label{d8} \\
&&\beta _{3}^{+}\left( \xi \right) =\Delta _{3}^{+-}\left( \xi \right)
=\Delta _{3}^{-+}\left( \xi \right) =\frac{\xi _{b}^{\ast }}{\left( \xi
_{b}^{\ast }+\xi \right) ^{2}};  \label{d9} \\
&&\Delta _{3}^{++}\left( \xi \right) =\frac{2\xi _{b}^{\ast 2}}{\left( \xi
_{b}^{\ast }+\xi \right) ^{3}}-\frac{\xi _{b}^{\ast }}{\left( \xi _{b}^{\ast
}+\xi \right) ^{2}},  \label{d10}
\end{eqnarray}%
where we introduced the notation
\[
\xi _{b}^{\ast }\equiv \frac{1}{\gamma _{b}}>0.
\]%
with the backscattering amplitude $\gamma _{b}$ defined in Eq.~(\ref{eq:4.40}%
). In Eqs. (\ref{d8}-\ref{d10}) one should write the final value of the $\xi
$ at which the renormalization stops. Its value can be written as
\begin{equation}
\xi \left( \theta ;u,u_{1};\mathbf{r}_{\perp }\right) =u_{1}u\mu _{d}\bar{f}%
\left( \mathbf{r}_{\perp }\right) \ln \left[ \min \left( \frac{1}{\theta },%
\frac{v_{F}}{r_{0}T}\right) \right]   \label{d2}
\end{equation}%
where $\theta $ is the angle between the vectors $\mathbf{n}$ and $\mathbf{n}%
_{1}$ and the function $\bar{f}$ is the Fourier transform of the function $%
f\left( k_{\parallel }=0,\mathbf{k}_{\perp }\right) $, Eq. (\ref{a7c}), with
respect to $\mathbf{k}_{\perp }$. The parameter $\mu _{d}$ equals: $\mu
_{1}=2;$ $\mu _{2}=4\left( p_{F}r_{0}\right) ^{-1};$ $\mu _{3}=4\pi \left(
p_{F}r_{0}\right) ^{-2}.$

One can see from Eqs. (\ref{d8}-\ref{d10}) that all the relevant vertices
decay as $\xi $ grows. This is usually referred to as the \textquotedblleft
zero charge\textquotedblright\ situation. If we start from small values of
the vertices (weak coupling) they become even smaller in the process of the
renormalization and the one loop approximation used here is sufficient for
writing the final results.

\section{Thermodynamic quantities.}

\label{thermodynamics}

\subsection{Specific heat.}

Calculation of thermodynamic quantities can be performed starting with a
standard relation for the thermodynamic potential $\Omega \left( T\right) $
\begin{equation}
\Omega \left( T\right) =-T\ln \mathcal{Z}  \label{f1}
\end{equation}%
where $\mathcal{Z}$ is the partition function.

We have performed the renormalization group calculations for the case when
the vectors $\mathbf{n}$ and $\mathbf{n}^{\prime }$ of two spin excitations
were close to each other (parallel or antiparallel motion). Only in this
limit one obtains large logarithms that determine the renormalization of the
vertices. A crucial question is whether or not this narrow region of the
parameters can bring an important contribution to thermodynamic or other
physical quantities. This is not a trivial question because the system was
not assumed to be one- or quasi-one-dimensional and one could imagine that
all the effect of the singularities in the vertices would be washed out
after the summation over the whole phase space.

In fact, this almost parallel motion of the spin excitations does not
contribute much into the thermodynamic potential $\Omega \left( T\right) $
itself. Fortunately, this is not a very interesting quantity and what one
would like to know are derivatives of the thermodynamic potential with
respect to temperature and other parameters. One of the most important
thermodynamic quantities is the specific heat $C$ that can be expressed
through the thermodynamic potential $\Omega $ as%
\begin{equation}
C=-T\frac{\partial ^{2}\Omega }{\partial T^{2}}  \label{f2}
\end{equation}

What we need is to calculate not the thermodynamic potential $\Omega\left(
T\right) $ itself but the difference $\delta\Omega\left( T\right) ,$
\begin{equation}
\delta\Omega\left( T\right) =\Omega\left( T\right) -\Omega\left( 0\right)
\label{f3}
\end{equation}
Using the diagrammatic method of the calculations we should be able to
express the thermodynamic potential $\Omega\left( T\right) $ in terms of
sums of products of the Green functions over bosonic Matsubara frequencies $%
\omega _{n}$
\begin{equation}
\Omega\left( T\right) =\sum_{\omega_{n}}R\left( \omega_{n}\right)
\label{f3a}
\end{equation}
where $R\left( \omega_{n}\right) $ is a function of the frequency.

The sums of the type of Eq. (\ref{f3a}) are very often divergent at high
frequencies. This problem can be avoided calculating the quantity $%
\delta\Omega\left( T\right) ,$ Eq. (\ref{f3}).

Using the Poisson formula we represent $\delta \Omega \left( T\right) $ in
the form%
\begin{equation}
\delta \Omega \left( T\right) =2\sum_{l=1}^{\infty }\int \frac{d\omega }{%
2\pi }R\left( \omega \right) \exp \left( -\frac{il\omega }{T}\right)
\label{f4}
\end{equation}%
which improves drastically the convergence. The essential frequencies in Eq.
(\ref{f4}) are of the order $T$ and are smaller then those frequencies that
form logarithms in the vertices.

We calculate the partition function $\mathcal{Z}_{t}$ using Eq. (\ref{c7}).
If we kept in the action $\mathcal{S}\left[ \psi \right] $ the
supersymmetric part $\mathcal{S}_{0}\left[ \psi \right] +\mathcal{S}_{4}%
\left[ \psi \right] $, Eqs. (\ref{c8}, \ref{c11}), only, we would get zero
for the thermodynamic potential $\Omega $. The terms $\mathcal{S}_{2}\left[
\psi \right] $ and $\mathcal{S}_{3}\left[ \psi \right] ,$ Eqs. (\ref{c9}, %
\ref{c10}), violate the supersymmetry and, as a result, one obtains finite
contributions to $\Omega $.

The thermodynamic potential $\Omega \left( T\right) $ can be expanded in
terms of the renormalized action $\mathcal{S}_{2}\left[ \psi \right] $ and
the lowest orders of the expansion take the form%
\begin{equation}
\Omega \left( T\right) =\Omega _{1}\left( T\right) +\Omega _{2}\left(
T\right) ,  \label{f5}
\end{equation}%
where
\begin{equation}
\Omega _{1}\left( T\right) =T\left\langle \mathcal{S}_{2}\left[ \psi \right]
\right\rangle _{0},\quad \Omega _{2}\left( T\right) =-\frac{T}{2}%
\left\langle \left( \mathcal{S}_{2}\left[ \psi \right] \right)
^{2}\right\rangle _{0}  \label{f6}
\end{equation}%
and $\left\langle ...\right\rangle _{0}$ means averaging over $\psi $ with
the Lagrangian $\mathcal{S}_{0}\left[ \psi \right] ,$ Eq. (\ref{c8}). The
quantities $\delta \Omega _{1,2}\left( T\right) $ are obtained for $\Omega
_{1,2}\left( T\right) $ by subtracting $\Omega _{1,2}\left( 0\right) .$

It turns out that the terms $\delta \Omega _{1}\left( T\right) $ and $\delta
\Omega _{2}\left( T\right) $ lead to qualitatively different types of
contributions. In the first order in the interaction, only the forward
scattering contributes. As we have found in the previous subsection, the
part of $\mathcal{S}_{2}$ corresponding to the forward scattering is not
renormalized and one may use just Eq. (\ref{c9}) for it. Then, we come to
the expression%
\begin{equation}
\frac{\delta C_{1}}{T}=\frac{\pi \left( 3\gamma _{f}\right) }{6v_{F}\lambda
_{0}^{d-1}}  \label{f7}
\end{equation}%
where $\lambda _{0}$ differs from $r_{0}$, Eq. (\ref{a7c}), by a numerical
coefficient. Actually, $\lambda _{0}$ should be of the order of the wave
length if one wants to make estimates for the initial model of the
interacting fermions. The factor $3$ in the numerator is due to the fact
that the spin-$1$ excitations have $3$ projections. A similar contribution
comes also from the density excitations but, of course, with the factor $1.$
Eq. (\ref{f7}) describes a contribution of the interaction to the
coefficient in front of the linear dependence of the specific heat on
temperature.

The part $\delta \Omega _{2}$ consists of the part further renormalizing the
coefficient $C/T$ of the linear dependence on $T$ and a part giving
corrections non-analytic in $T^{2}$. A general expression for $\delta \Omega
_{2}\left( T\right) $ can be brought to the form
\begin{eqnarray}
&&\delta \Omega _{2}\left( T\right) =-6\lim_{\eta \rightarrow +0}\
\sum_{l=1}^{\infty }\int \frac{d\omega }{(2\pi )}\exp \left( -i\frac{l\omega
}{T}-\eta \left\vert \omega \right\vert \right)  \nonumber \\
&&\times \int \!\!d\mathbf{n}_{1}d\mathbf{n}_{2}\int \frac{d^{d}\mathbf{k}}{%
\left( 2\pi \right) ^{d}}Y\left( \theta ;\mathbf{k}_{\perp },k_{\parallel
}\right) \mathcal{P}_{d}\left( \omega ,\mathbf{k;n}_{1},\mathbf{n}_{2}\right)
\label{f12}
\end{eqnarray}%
where $\theta $ is the angle between the vectors $\mathbf{n}_{1}$ and $%
\mathbf{n}_{2}.$ The main interesting contribution will come from small $%
\theta $, which justifies the decomposition of the momentum $\mathbf{k}$
into perpendicular $\mathbf{k}_{\perp }$ and parallel $k_{\parallel }$ with
respect to $\mathbf{n}_{1,2}$ components.

The function $Y\left( \theta ;\mathbf{k}_{\perp },k_{\parallel }\right) $
defined as
\begin{eqnarray}
&&Y\left( \theta ;\mathbf{k}_{\perp },k_{\parallel }\right)
=\int_{0}^{1}\int_{0}^{1}u_{1}u_{2}du_{1}du_{2}  \label{f13} \\
&&\times \left\{ \left[ \Delta _{3}^{+-}\left( \theta ;u_{1},u_{2};\mathbf{k}%
_{\perp },k_{\parallel }\right) \right] ^{2}+\Delta _{3}^{++}\left( \theta
;u_{1},u_{2};\mathbf{k}_{\perp },k_{\parallel }\right) \Delta
_{3}^{--}\left( \theta ;u_{1},u_{2};\mathbf{k}_{\perp },k_{\parallel
}\right) \right\}  \nonumber
\end{eqnarray}%
is the most important entry in the final expression for the specific heat.
The vertices $\Delta _{3}$ should be taken from Eqs. (\ref{d9}-\ref{d2}).

The formfactor
\begin{equation}
\mathcal{P}_{d}\left( \omega ,\mathbf{n}{_{1},\mathbf{n}_{2}}\right) =\frac{%
\left( i\omega +v_{F}\mathbf{n}_{2}\right) \left( i\omega -v_{F}\mathbf{n}%
_{1}\right) }{\left( i\omega -v_{F}\mathbf{n}_{2}\ \right) \left( i\omega
+v_{F}\mathbf{n}_{1}\right) }  \label{f14}
\end{equation}%
depends on the dimensionality of the system and it describes basically the
free propagation of the two spin excitations in almost opposite directions.
The non-analytic contributions originate from the small region of the phase
space $\left\vert \mathbf{n}_{1}-\mathbf{n}_{2}\right\vert \ll 1$.

The result of the calculation depends on the dimensionality of the system
but the non-trivial corrections exist in both $d=2$ and $d=3$. The details
of the calculations can be found in Ref. \cite{aleiner} but the final
results for the non-analytic corrections $\delta C$ to the specific heat
look for $d=2$ and $d=3$ as%
\begin{equation}
\delta C_{d=2}=-\frac{3\zeta \left( 3\right) T^{2}}{\pi \varepsilon _{F}^{2}}%
\left\{ \left[ \gamma _{b}^{\rho }\right] ^{2}+\frac{12\gamma _{b}^{2}\ln
^{2}\left[ 1+\mathcal{X}\left( T\right) /2\right] }{\mathcal{X}^{2}\left(
T\right) }\right\}  \label{f15}
\end{equation}%
\begin{equation}
\delta C_{d=3}=-\frac{3\pi ^{4}}{10}\left( \frac{T}{\varepsilon _{F}}\right)
^{3}\left\{ \left[ \gamma _{b}^{\rho }\right] ^{2}\ln \frac{\varepsilon _{F}%
}{T}+\frac{6\pi \gamma _{b}}{\mu _{3}}\int_{0}^{\frac{\mathcal{X}\left(
T\right) }{2\pi }}\frac{dz}{z^{2}}\left[ \mathrm{Li}_{\mathrm{2}}\left(
-z\right) \right] ^{2}\right\}  \label{f16}
\end{equation}%
where $\mathcal{X}\left( T\right) =\mu _{d}\gamma _{b}\ln \left( \varepsilon
_{0}/T\right) $ ($\varepsilon _{0}=v_{F}/r_{0}\simeq \varepsilon _{F}$ and $%
\mathrm{Li}_{\mathrm{2}}\left( x\right) =\sum_{k=1}^{\infty }x^{k}/k^{2}$ is
the polylogarithm function. The first terms in the circular brackets in Eqs.
(\ref{f15}, \ref{f16}) describe the density excitations and $\gamma
_{b}^{\rho }$ is the coupling constant for the backward scattering of these
excitations. The second terms originate from the spin excitations. The final
result, Eqs. (\ref{f15}, \ref{f16}), was written for a special choice of the
cutoff function $\bar{f}\left( \mathbf{r}_{\perp }\right) $, Eq. (\ref{d2})
\begin{equation}
\bar{f}\left( \mathbf{r}_{\perp }\right) =\Omega _{d}^{-1}\exp \left(
-r_{\perp }/r_{0}\right)  \label{f16a}
\end{equation}%
where $\Omega _{d}$ is the $d-1$ dimensional solid angle ($\Omega
_{1}=2,\Omega _{2}=2\pi $).

In the limit of not very low temperatures when $\mathcal{X}\left( T\right)
\ll 1$, Eqs. (\ref{f15}, \ref{f16}) take a simpler form%
\begin{equation}
\delta C_{d=2}=-\frac{3\zeta \left( 3\right) }{\pi }\left( \frac{T}{%
\varepsilon _{F}}\right) ^{2}\left( \left[ \gamma _{b}^{\rho }\right]
^{2}+3\gamma _{b}^{2}\right)  \label{f17}
\end{equation}%
\begin{equation}
\delta C_{d=3}=-\frac{3\pi ^{4}}{10}\left( \frac{T}{\varepsilon _{F}}\right)
^{3}\ln \left( \frac{\varepsilon _{F}}{T}\right) \left( \left[ \gamma
_{b}^{\rho }\right] ^{2}+3\gamma _{b}^{2}\right)  \label{f18}
\end{equation}%
Eqs. (\ref{f17}, \ref{f18}) agree with results obtained by a direct
diagrammatic expansions (see for a recent discussion Refs. \cite%
{chubukov3,chubukov4}. At the same time, the general equations (\ref{f5}, %
\ref{f6}) have been obtained for the first time in Ref. \cite{aleiner} using
the bosonization scheme. We see that in the limit $T\rightarrow 0$ the
contribution of the spin excitations vanishes and only the density
excitations contribute to the specific heat.

The bosonization method presented here gives a possibility to make
calculations also for one-dimensional systems. Surprisingly, the calculation
of the specific heat $1d$ is more complicated than for $d=2,3$. This is
because the function $\mathcal{P}_{d}\left( \omega ,\mathbf{n}{_{1},\mathbf{n%
}_{2}}\right) $, Eq. (\ref{f14}), is exactly equal to unity and the
contribution of Eq. (\ref{f12}) to the specific heat vanishes. In order to
obtain a non-vanishing contribution one should consider terms of higher
orders in the effective vertices. As a result, one comes to the following
expression for the correction to the specific heat in $d=1$%
\begin{equation}
\delta C_{d=1}=\frac{\pi T}{v_{F}}\frac{1}{\left( 1+2\gamma _{b}\ln \frac{%
\varepsilon _{F}}{T}\right) ^{3}}  \label{f19}
\end{equation}%
The correction to the specific heat for $1d$ can be extracted from the exact
solution for spin chains of Ref. \cite{lukyanov}. This correction agrees
with our result, Eq. (\ref{f19}), and we see that our supersymmetric low
energy theory reproduces all the previously known physical effects despite
the fact that the intermediate degrees of freedom differ from the
conventional bosonization.

\subsection{Spin susceptibility}

Calculation of the spin susceptibility has been performed in a recent work
\cite{schwiete}. The bosonization scheme of Ref. \cite{aleiner} was used and
it was shown that the logarithmic contributions arise as well. An external
magnetic field \ $\mathbf{b}$ leads to the additional terms $\mathcal{S}%
_{b0},\mathcal{S}_{b1}$ and $\mathcal{S}_{b2}$, Eqs. (\ref{c16}-\ref{c17}),
in the effective action.

We proceed as before integrating over the fast variables $\Phi $, Eq. (\ref%
{c40}) and thus deriving renormalization group equations. In principle, one
should renormalize not only the terms $\mathcal{S}_{2},\mathcal{S}_{3},%
\mathcal{S}_{4}$, Eqs. (\ref{c9}-\ref{c11}), as it was done previously, but
also the terms $\mathcal{S}_{b0},\mathcal{S}_{b1}$ and $\mathcal{S}_{b2},$
Eqs. (\ref{c16}-\ref{c17}) and $\mathcal{S}_{0},$ Eq. (\ref{c8}). Again, the
calculations should be done separately for \ $d=1$ and $d=2,3.$

It turns out that in the higher dimensions $d=2,3$ the terms $\mathcal{S}%
_{b0},\mathcal{S}_{b1}$ and $\mathcal{S}_{b2},$ and $\mathcal{S}_{0}$ do not
change under the renormalization and what remains to be done is to express
the susceptibility $\chi $ in terms of the renormalized vertices $\gamma
\left( \xi \right) $, $\beta \left( \xi \right) $ and $\Delta \left( \xi
\right) $. The computation is somewhat more cumbersome than the one for the
specific heat and there are many contributions that have to be combined
together. As for the specific heat, only the vertices with $\ i=3$
contribute and one can express the susceptibility through $\gamma _{3}\left(
\xi \right) $, $\beta _{3}\left( \xi \right) $ and $\Delta _{3}\left( \xi
\right) $.

Using the solutions of the RG equations, Eqs. (\ref{d8}-\ref{d10}), one can
bring the expression for the susceptibility to a form containing the
function $Y\left( \theta ;\mathbf{k}_{\perp },k_{\parallel }\right) $, Eq. (%
\ref{f13}). Choosing again the cutoff function $\bar{f}$ in the form of Eq. (%
\ref{f16a}) we bring the temperature dependent correction $\delta \chi
\left( T\right) $ to the spin susceptibility to the form \cite{schwiete}%
\begin{equation}
\delta \chi _{d=2}\left( T\right) =8\eta ^{2}\gamma _{b}^{2}\frac{T}{%
\varepsilon _{F}}\chi _{d=2}^{\left( 0\right) }\frac{\ln ^{2}\left[ 1+%
\mathcal{X}\left( T\right) /2\right] }{\mathcal{X}^{2}\left( T\right) }
\label{f20}
\end{equation}%
\begin{equation}
\delta \chi _{d=3}\left( T\right) =\frac{\left( 2\pi ^{2}\eta \right) ^{2}}{3%
}\gamma _{b}^{2}\left( \frac{T}{\varepsilon _{F}}\right) ^{2}\chi
_{d=3}^{\left( 0\right) }\frac{\left\{ \mathrm{Li}_{\mathrm{2}}\left[ -%
\mathcal{X}\left( T\right) /2\pi \right] \right\} ^{2}}{\mathcal{X}%
^{2}\left( T\right) }  \label{f21}
\end{equation}%
where $\eta $ is given by Eq. (\ref{eq:4.48}). The susceptibility $\chi
_{d}^{\left( 0\right) }$ is the Pauli susceptibility in $d$ dimensions.

In the limit $\mathcal{X}\left( T\right) \ll 1$ one can neglect the
logarithmic contributions and obtain the following form of the corrections $%
\delta \chi _{d}$%
\begin{equation}
\delta \chi _{d=2}\left( T\right) =2\eta ^{2}\gamma _{b}^{2}\frac{T}{%
\varepsilon _{F}}\chi _{d=2}^{\left( 0\right) }  \label{f22}
\end{equation}%
\begin{equation}
\delta \chi _{d=3}\left( T\right) =\frac{\pi ^{2}}{3}\eta ^{2}\gamma
_{b}^{2}\left( \frac{T}{\varepsilon _{F}}\right) ^{2}\chi _{d=3}^{\left(
0\right) }  \label{f23}
\end{equation}%
The linear dependence on $T$ of the correction $\chi \left( T\right) $, Eq. (%
\ref{f22}), agrees with those obtained by conventional methods \cite{coffey,
chubukov1,sarma,catelani}. At the same time, the proportionality of $\delta
\chi \left( T\right) $ to $T^{2}$ in $d=3$ is analytical in $T^{2}$ and Eq. (%
\ref{f23}) describes a renormalization of a coefficient in front of the $%
T^{2}$. As the $T^{2}$ -term is present already in the temperature
dependence of the susceptibility of the ideal Fermi gas, the correction is
not very interesting. The first non-analytical term $T^{2}\ln \left(
\varepsilon _{F}/T\right) $ in the temperature dependence in $d=3$ is
proportional to $\gamma _{b}^{3}$.

Calculation of the susceptibility for one dimensional systems is somewhat
more involved because on should consider corrections to the terms $\mathcal{S%
}_{b0},\mathcal{S}_{b1}$ and $\mathcal{S}_{b2},$ Eqs. (\ref{c16}-\ref{c17})
and $\mathcal{S}_{0},$ Eq. (\ref{c8}). Nevertheless, one can proceed in a
rather straightforward way and, as a results, the following dependence of
the susceptibility is obtained \cite{schwiete}%
\begin{equation}
\delta \chi _{d=1}\left( T\right) =\frac{2\nu \gamma _{b}}{1+2\gamma _{b}\ln
\left( \varepsilon _{F}/T\right) }  \label{f24}
\end{equation}%
This result agrees the with the one obtained long ago \cite{dl} by a
completely different RG method developed for the electron problem in $1d$.
So, Eq. (\ref{f24}) serves as one more check of the bosonization method
reviewed here.

\section{Discussion.}

\label{discussion}

In the previous sections we reviewed the new method of bosonization for a
clean Fermi gas in any dimensions suggested in Ref. \cite{aleiner} and
further used in Ref. \cite{schwiete}. In contrast to previous attempts \cite%
{haldane,houghton1,houghton,neto,kopietz,kopietzb,khveshchenko,khveshchenko1,metzner}
we do not restrict ourselves with the case of a long range electron-electron
interaction and include into the scheme spin degrees of freedom. This
enables us to consider not only density excitations but also the spin ones.
In contrast to the density excitations that can be described by a scalar,
the spin excitations are described by $3$-component vectors and one may
speak about a new version of the non-abelian bosonization.

The non-abelian character of the effective theory leads to the non-trivial
interaction between the spin modes. Making the perturbation theory in this
interaction new logarithmic contributions (diverging in the limit $%
T\rightarrow 0$) to vertices were discovered and summed up using a RG
scheme. As a result of this consideration we have found temperature
dependent corrections $\delta C\left( T\right) /T$ to the specific heat and
to the susceptibility $\delta \chi \left( T\right) $ in all dimensions, Eqs.
(\ref{f15}, \ref{f16}, \ref{f19}, \ref{f20}, \ref{f21}). The results for $%
d=1 $ agree with those obtained earlier by completely different methods. In
higher dimensions, $d=2,3$ the lowest order of the expansion of our formulae
in the coupling constant agree with known results obtained previously using
conventional diagrammatic techniques, see e.g. \cite%
{chubukov3,chubukov4,coffey, chubukov1,sarma,catelani}. All these agreements
serve as a good check of our approach.

The new contributions to the specific heat and susceptibility originate from
an almost parallel motion of the spin excitations. Although both the forward
and backward scattering amplitudes are renormalized, only the backward
scattering enters the thermodynamic quantities.

In the language of conventional diagrams the logarithmic contributions to
the thermodynamic quantities come from both Cooper and particle-hole loops
because they originate from quasi-one-dimensional processes. The fact that
the forward scattering amplitude drops out from the final results
corresponds to what happens in $1d$.

Although we have agreement in the limiting cases with almost all
the results we could compare with, there is a disagreement with an
old work \cite{kl}, where an instability of the Fermi liquid
against superconducting pairing with a high angular momentum was
found. This is especially strange because the main contribution to
the formation of the superconductivity comes from almost parallel
electron motion such that only forward and backward scattering
enter the superconducting critical temperature. This is just the
region that we considered. As we do not see any such effect, this
can mean that either 1) the accuracy of our method is not
sufficient (we summed the terms like $\left( \gamma \ln \left(
\varepsilon _{F}/T\right) \right) ^{n},$ whereas $\left( \gamma
^{2}\ln \left( \varepsilon _{F}/T\right) \right) ^{n}$ were summed
in Ref. \cite{kl}) or 2) summing the Cooper ladder as was done in
Ref. \cite{kl} is a bad approximation for the
quasi-one-dimensional process. We believe that 2) is the
explanation of the contradiction but a more careful analysis
should be performed.

We have developed and applied the bosonization scheme for the simplest model
of the interacting Fermi gas. One can add other terms to the Hamiltonian of
this model to take into account different features relevant for experimental
systems. We hope that our scheme may be useful in this study.

\end{document}